\documentclass[cpp,fleqn]{w-art}
\usepackage{times}
\usepackage{w-thm} 
\usepackage{multirow}
\theoremstyle{plain}

\theoremstyle{definition}

\usepackage[]{graphicx}
\chardef\bslash=`\\ 

\begin{document}
\DOIsuffix{theDOIsuffix}
\Volume{42}
\Issue{1}
\Month{01}
\Year{2003}
\pagespan{3}{}
\Receiveddate{03 November 2006}
\Reviseddate{}
\Accepteddate{}
\Dateposted{}
\keywords{density functional molecular dynamics, equation of state, giant gas planets, hydrogen-helium mixtures}
\subjclass[pacs]{61.20.Ja, 61.25.Em, 61.25.Mv, 61.20.-p}



\title[Dense Fluid Hydrogen and Helium]{Properties of Dense Fluid Hydrogen 
and Helium in Giant Gas Planets}


\author[Vorberger]{Jan Vorberger\footnote{Corresponding
     author: e-mail: {\sf jvorberger@ciw.edu}, Phone: +001\,202\,478\,8943,
     Fax: +001\,202\,478\,8901}\inst{1}} \address[\inst{1}]{Geophysical 
     Laboratory, Carnegie Institution of Washington, Washington D.C., USA}
\author[Tamblyn]{Isaac Tamblyn\inst{2}}
\address[\inst{2}]{Department of Physics, Dalhousie University, Halifax, Nova Scotia, Canada}
\author[Bonev]{Stanimir A. Bonev\inst{2}}
\author[Militzer]{Burkhard Militzer\inst{1}}
\begin{abstract}
Equilibrium properties of hydrogen-helium mixtures under thermodynamic conditions 
found in the interior of giant gas planets are studied by means of density
functional theory molecular dynamics simulations. Special emphasis is placed on the 
molecular-to-atomic transition in the fluid phase of hydrogen in
the presence of helium.
Helium has a substantial influence on the stability of hydrogen molecules. The molecular bond
is strengthened and its length is shortened as a result of the increased 
localization of
the electron charge around the helium atoms, which leads to more stable hydrogen molecules 
compared to pure hydrogen for the same thermodynamic conditions. The
{\it ab initio} 
treatment of the mixture enables us to investigate the structure of the liquid
and to discuss hydrogen-hydrogen, helium-helium, and hydrogen-helium
correlations on the basis of pair correlation functions.
\end{abstract}
\maketitle                   





\section{Introduction}\label{intr}
The discovery of the first extrasolar planet in 1995~\cite{mayor95} 
marked the beginning of a new era in
planetary science, characterized by a rapidly expanding 
set of known extrasolar planets. More than 200 exoplanets have been discovered
so far \cite{exopl}. 
Among these, giant gas planets in small orbits are in the majority since the primary tool for 
detection, radio velocity measurements, is most sensitive to finding 
heavy planets that rapidly orbit their parent star \cite{Bu05,rivera05}.

From radius measurements of transient extrasolar planets, it is also known that 
most of these giant gas planets
are like Jupiter in consisting primarily of hydrogen and helium.
Modeling the interior of such planets requires an accurate equation of state for
hydrogen-helium mixtures at high pressure and temperature conditions similar to 
those in planetary interiors \cite{Gu02}. Thus, the characterization of such 
system by
first principle calculations will help us to answer questions concerning the
inner structure of planets, their origin and evolution \cite{Gu02,SG04}.

In this article, we focus on studying the transition from molecular to atomic 
hydrogen. 
In particular, we investigate the effect of different helium concentrations 
on this transition. 
In what follows, it will be shown how the structure of the fluid and its 
equation of state (EOS) change with varying the hydrogen-helium mixing ratio.
\section{Method}
\begin{figure}[ht]
\includegraphics[width=.45\textwidth]{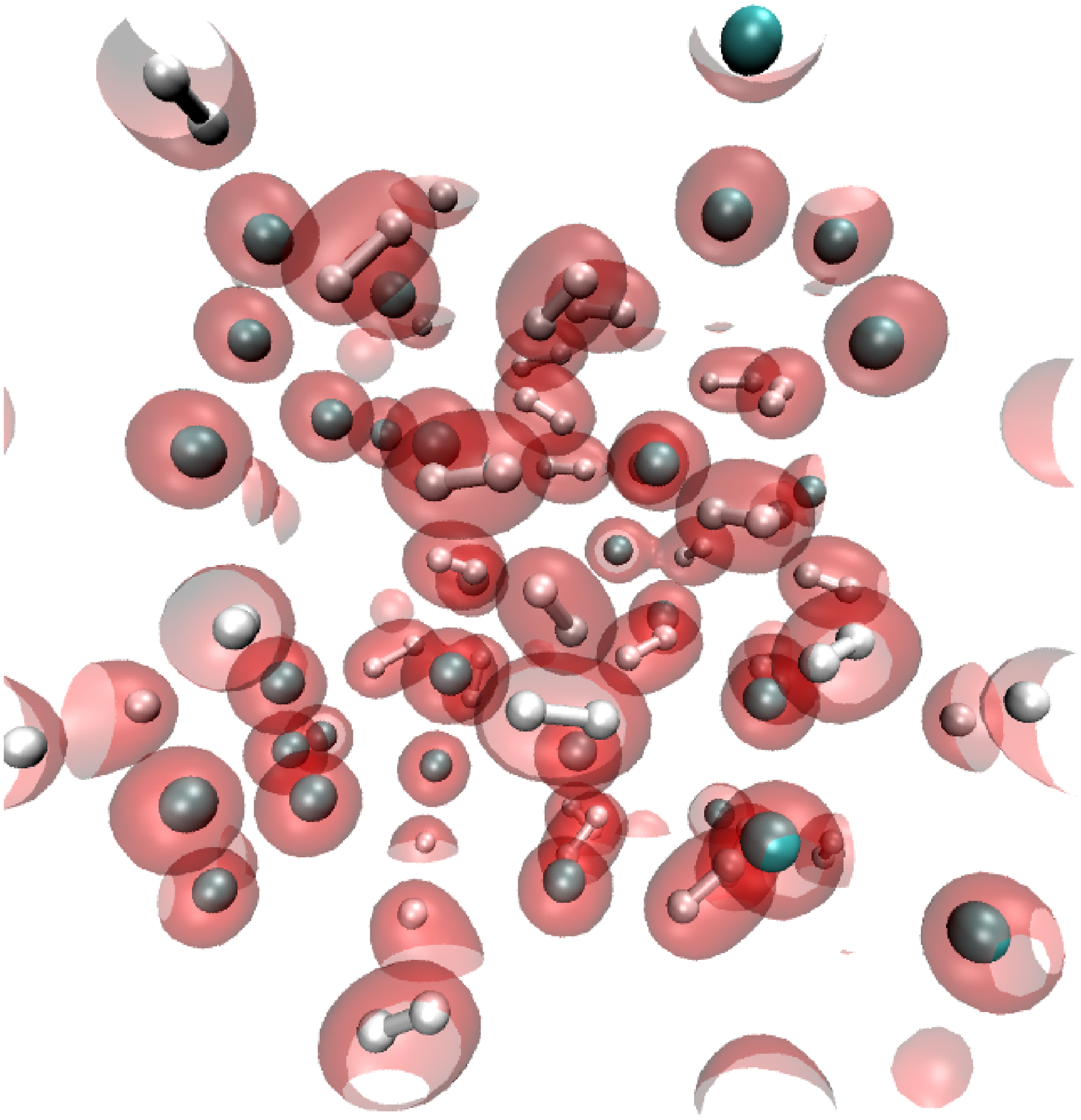}~a)
\hfil
\includegraphics[width=.45\textwidth]{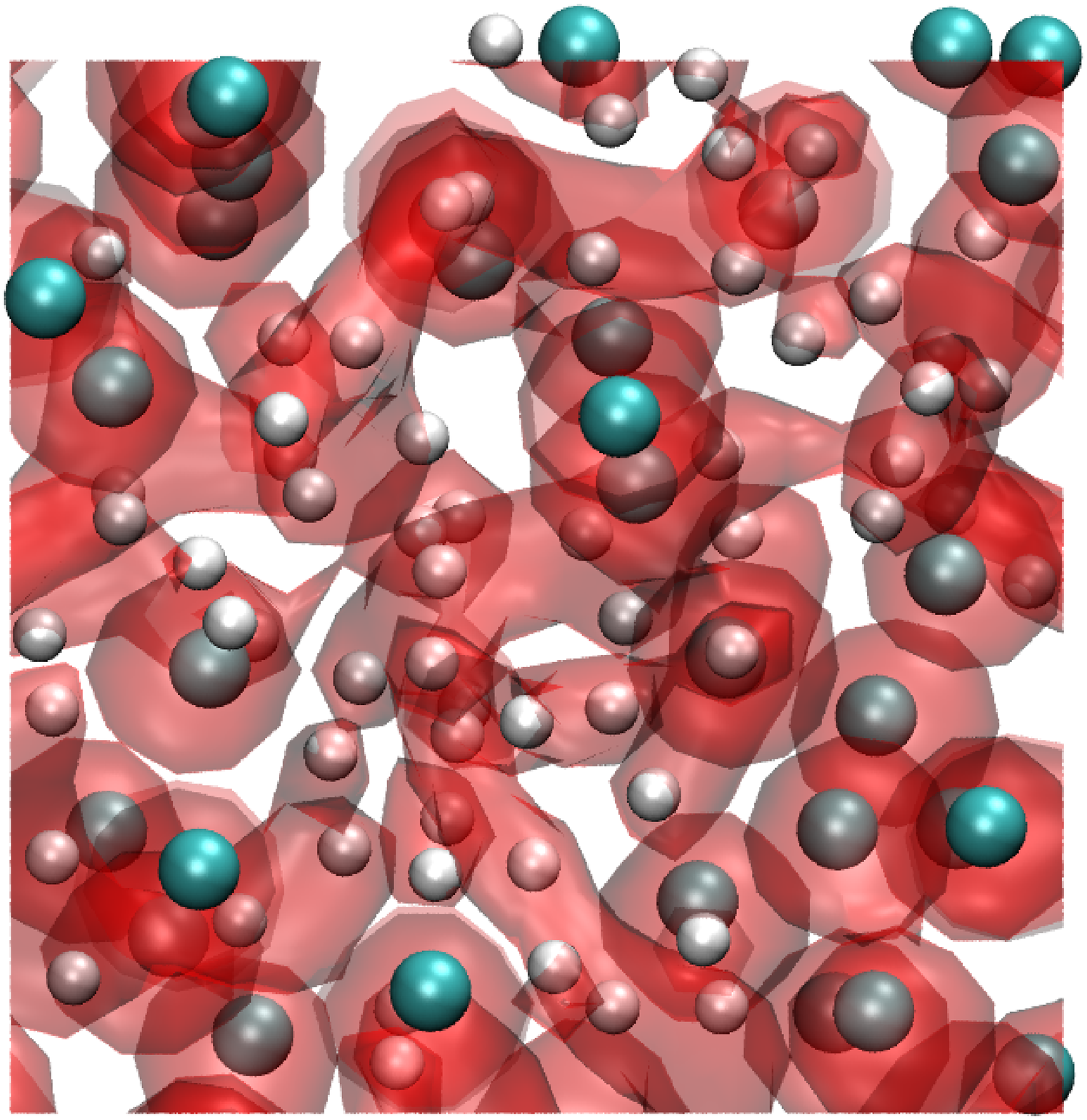}~b)
\caption{View of the simulation cell of a DFT-MD simulation. Hydrogen and 
helium nuclei are represented by small and large spheres, respectively, 
and electron density 
isosurfaces are drawn. In these plots, the mixing ratio is $x=0.5$,
and the temperature is $4000$ K. 
The density corresponds to a) $r_s=2.4$ ($0.3$ g/cm$^3$) at pressure of  
$8$ GPa, and b) $r_s=0.9$ ($5.5$ g/cm$^3$) 
at pressure of $3800$ GPa. 
These conditions correspond to a molecular phase and to a metallic regime
of fluid hydrogen, respectively.
}
\label{fig:nice}
\end{figure}
We use density functional molecular dynamics (DFT-MD) to investigate the 
questions connected with the physics of giant gas planets.
Under conditions along Jupiter's isentrope, the electrons can be considered 
to be in their ground state. They are 
either chemically bound in molecules or atoms, or form a degenerate interacting 
electron gas. Density functional theory can thus be used to describe the 
electrons throughout.
The ions form a strongly coupled fluid which makes many-body
simulations an obvious choice for 
their description. For the interaction of the classical ions 
and electrons (quantum particles) the 
Born-Oppenheimer (BO) approximation can be invoked since the motions of electrons
and ions follow different timescales.

The results presented here are obtained using the CPMD code \cite{CPMD}. 
We use simulations cells with $128$ electrons and the corresponding number 
of ions, and periodic boundary conditions.
The forces acting on the ions are provided by DFT calculations within
generalized gradient approximation (GGA) 
for the electrons in the Coulomb field of the ions. We use 
Troullier Martin norm-conserving 
pseudopotentials for the electron-ion interactions \cite{TM91}.
To check for finite-size effects, calculations with supercells ranging from 
54 to 250 atoms are carried out, but give no significant change in the results
($<2\%$ in pressure for supercells with 128 or more particles). 
We have also performed simulations with different number of k points (1 to 512) 
to sample the Brillouin zone, and it was confirmed that a single k point already
gives converged results for the supercells and conditions used here.
Recently, it was shown that in high pressure water effects arising from the 
treatment of the electrons at their true temperature instead of in the 
ground state can be important for the dynamics of the system \cite{Mattson06}. 
Therefore, special attention was given to the effects arising from a finite 
electron temperature. 
We have performed additional DFT-MD simulations with the VASP code
using the finite temperature 
Mermin functional to model the influence of thermal occupation of electronic 
states \cite{VASP}. For Jupiter's mixing ratio 
of $x=N_{He}/(2N_{H}+N_{He})=0.14$ almost no difference could be detected at 
relevant temperatures in the metallic regime (less than half a percent 
deviation).
%
%
\section{Equation of State and Structure of the Hydrogen-Helium Fluid}
Figure \ref{fig:nice} displays snapshots of the simulation cell during runs for
two different 
situations. In Fig. \ref{fig:nice}a) the molecular fluid at low density 
($r_s=2.4$) is characterized 
by larger distances between the molecules and by electronic densities concentrated
around the nuclei. Such a molecular fluid can be found 
in the outer mantle regions of Jupiter. The physical system reminiscent to the 
inner mantle of Jupiter, shown in Fig. \ref{fig:nice}b), can be characterized as
a metallic fluid. The density is much higher, 
molecules are dissociated as a result of the increased pressure. 
The electrons are delocalized as a result of the Pauli exclusion principle and
form an electron cloud that extends through the simulation cell.

\begin{figure}[h]
\includegraphics[width=1.0\textwidth]{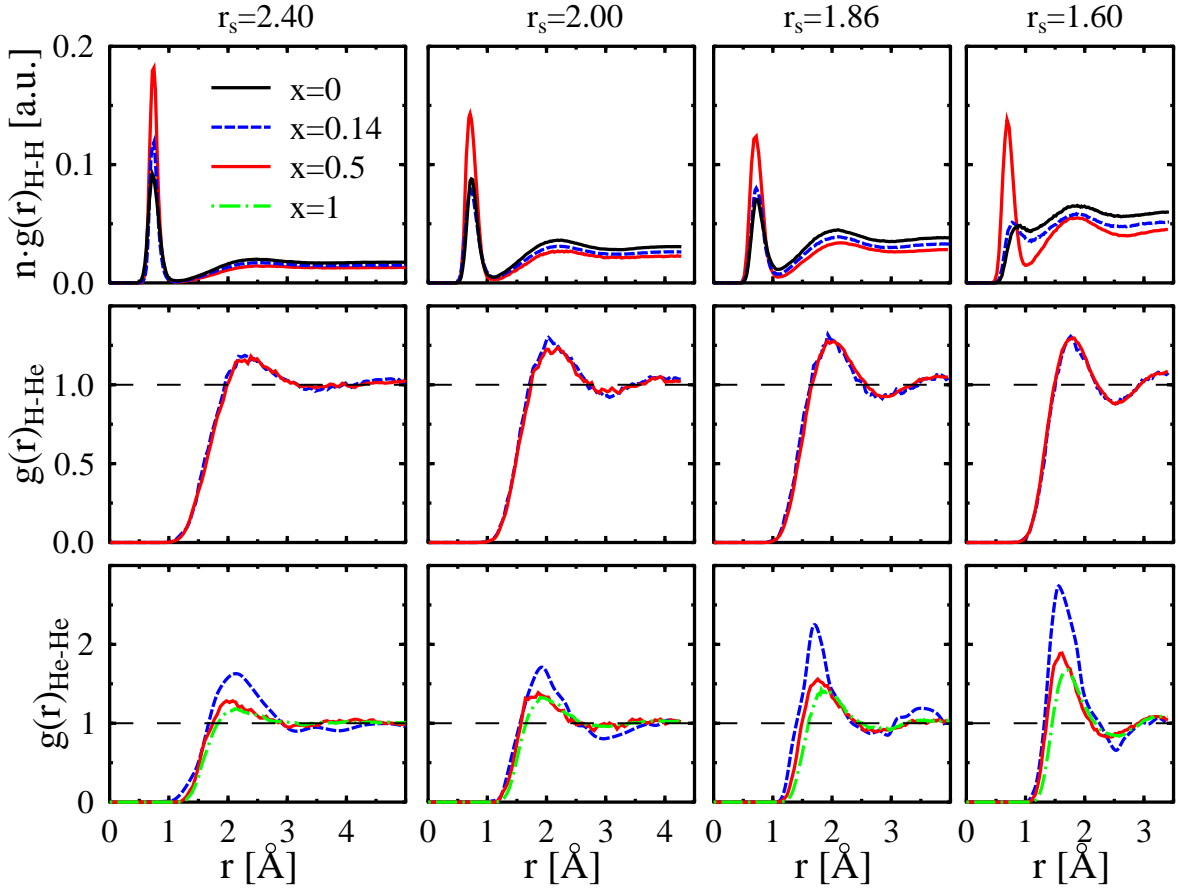}
\caption{
Pair correlation functions at a temperature of $3500$ K across the 
molecular-atomic transition for various densities and mixing ratios. 
The three rows of graphs show the hydrogen-hydrogen, hydrogen-helium, and
helium-helium pair correlation functions. Here, $g(r)_{H-H}$ was multiplied by the
concentration of hydrogen atoms so that the area under the peak at 
$r=0.75\;\mbox{A}$
corresponds to the fraction of molecules in the fluid. In each graph, simulations
for different mixing ratios have been combined: pure hydrogen ($x=0$), hydrogen
and helium for Jupiter's mixing ratio ($x=0.14$), a fluid with as many helium
atoms as hydrogen molecules ($x=0.5$), and pure helium ($x=1$). The columns show
results from different electronic densities (given in terms of the Wigner Seitz
parameter $r_s$), for which different mixing ratios
have been compared.
%
}
\label{fig:grarray}
\end{figure}
For given temperature and density conditions, which of the two described phases 
can be found, namely, molecular or atomic/metallic, depends strongly on 
the helium concentration. Moreover, the helium fraction
plays a crucial role in how the mixture transforms from the molecular 
to the atomic phase with increasing temperature or pressure.

Changes in the structure of the fluid, i.e., the transition from a molecular to 
a metallic fluid may be discussed in terms of the pair correlation function (PCF)
\begin{equation}
g(r)=\frac{V}{N(N-1)}\left\langle \sum_{i\neq j}\delta(r-|r_{ij}|)
\right\rangle\frac{1}{4\pi r^2}\;.
\end{equation}
Pair correlation functions for different types of nuclei are shown in Fig.
\ref{fig:grarray}.
We discuss the changes in the pair correlations with respect to 
density at a constant temperature of $3500$ K.
Let us first comment on the hydrogen-hydrogen pair correlations in pure hydrogen.
A pronounced peak at about $r=0.75$ \AA\, demonstrates the presence of molecules 
in the fluid. These pair correlations are normalized so that the area under
the molecular peak gives the number of molecules found in the fluid. The
dissociation degree can thus be extracted either by integrating over the PCF or
by invoking a distance criterion for two hydrogen atoms.  
Table \ref{valuetable} reports degrees of dissociation that where derived by
analyzing the MD trajectories using the distance criterion and a minimum
molecular lifetime of $10$ vibrational periods of the isolated hydrogen molecule.
The lifetime criteria leads to a decaying molecular fraction in the metallic
regime, which would be overestimated based on the distance criterion alone.
 
A high peak followed by a minimum that 
approaches zero is a strong indication that the molecules are stable, 
for instance this is the case at $r_s=2.4$ for different values of $x$. 
Indeed, this can be confirmed
by directly analyzing the number and lifetime of the molecules \cite{Vorberger06}.
On the other hand,
at high density ($r_s=1.6$, $x=0$), no isolated first peak can 
be identified and the shape of the PCF is 
typical for a correlated atomic liquid.
The first row of Fig. \ref{fig:grarray} contains two additional graphs showing 
the pair correlations at densities and mixing ratios intermediate between the 
two extrema just discussed. 
In these cases, the molecular peak can still be identified but its height is 
reduced and it is widened so that the first minimum has a value greater than 
zero. A shape like this is caused by a mixture of 
atoms and molecules; where the definition of molecules here is somewhat 
ambiguous as they may have finite lifetimes. 

Let us now discuss the influence of helium on the PCF. 
The general effect can be seen by comparing hydrogen-hydrogen pair correlation 
functions at identical electronic densities and different helium fractions.
More helium content in the mixture generally leads to a lower hydrogen 
dissociation. 
The presence of helium atoms thus stabilizes the hydrogen molecules and
the effect of pressure to dissociate hydrogen molecules is partially
suppressed. 
A more detailed analysis revealed that in mixtures of hydrogen and helium 
the bond length of the hydrogen molecule is 
shortened by about $6\%$ \cite{Pfaffenzeller1995,Vorberger06}. The influence of 
helium allows hydrogen molecules to exist in a parameter regime where 
otherwise pure hydrogen would be dissociated. The microscopic reason for 
this effect is the double charge of the 
helium nuclei binding two electrons. These electrons are bound 
stronger than the electrons in a hydrogen molecule and are more
localized, which has profound effects on the high pressure properties of the 
fluid \cite{Militzer06}. In the mixture, atoms and molecules can be packed 
more closely before electronic wavefunctions begin to overlap.
When molecules are compressed and wave functions begin 
to overlap, their localization increases because of the exclusion principle, 
which means larger kinetic energy. But the kinetic and potential energies 
scale differently with density, so eventually, it is energetically more 
favorable for the electrons to delocalize (lower kinetic energy, higher 
potential energy).

We also examine helium-helium (He-He) and hydrogen-helium (H-He) 
pair correlation functions
extracted from the DFT-MD simulations. Both, He-He and H-He pair 
correlation functions show a finite probability for finding a 
neighboring atom only for distances larger than $1.5$ \AA. 
This value reflects the bigger radius of a helium atom compared to
a hydrogen atom.

Hydrogen-helium correlations are similar to the continuous part of the 
hydrogen-hydrogen PCFs. Naturally, there is no molecular peak and significant
contributions occur at distances larger than the first minimum of the 
hydrogen-hydrogen PCFs. Thereafter, similar weak short-range 
order expressed in oscillations around unity can be observed. There are only
minor differences between hydrogen-helium correlation functions of
different mixing ratios.

\begin{table}[t]
\begin{center}
\begin{tabular}{c||c|c|c|c}
&$r_s=2.40$&$r_s=2.00$&$r_s=1.86$&$r_s=1.60$\\\hline\hline
\multirow{2}{*}{x=0.0}
&$8.9 \pm0.1$ GPa &$25.5\pm 0.1$ GPa&$38.5\pm
0.3$ GPa&$89.4\pm 0.2$ GPa\\
&$0.94$&$0.38$&$0.20$&$0.05$\\\hline
\multirow{2}{*}{x=0.14}
&$8.5\pm 0.1$ GPa&$24.1\pm 0.2$ GPa&$38.4\pm 0.2$ GPa&$90.8\pm 0.3$ GPa\\
&$0.98$&$0.38$&$0.43$&$0.09$\\\hline
\multirow{2}{*}{x=0.5}
&$7.0\pm 0.1$ GPa&$21.2\pm 0.1$ GPa&$33.0\pm 0.2$ GPa&$89.7\pm 0.2$ GPa\\
&$0.97$&$0.45$&$0.44$&$0.49$\\\hline
\multirow{2}{*}{x=1.0}
&$5.3\pm 0.1$ GPa&$14.7\pm 0.1$ GPa&$22.5\pm 0.1$ GPa&$65.0\pm 0.1$ GPa \\
&--&--&--&--\\
\end{tabular}
\caption{
Values for pressure (1st line) and fraction of atoms bound in molecules 
(2nd line) at various densities and mixing
ratios at $T=3500$ K.
 These values belong to the pair correlation functions in
Fig. \ref{fig:grarray}.
}
\label{valuetable}
\end{center}
\end{table}
The correlations in the helium subsystem can be
described as short-range hard sphere type interactions. Interestingly, the
helium-helium pair correlations show a more pronounced
first peak in the $x=0.14$ mixture than in the $x=0.5$ mixture 
and they again feature a slightly higher peak than the ones for pure helium. 
There is a bigger total number of hydrogen atoms 
present for a lower mixing ratio $x$ at constant electronic density. 
Two hydrogen atoms or one hydrogen molecule have a bigger volume than one helium
atom. The effective density for the helium atoms is thus higher in a mixture 
with less helium. This causes a general requirement to increased coordination 
among the atoms which in turn is reflected in the higher peak in the pair 
correlation. The effect of hydrogen on helium can be understood as the 
opposite of the effect of helium on hydrogen.

The observed structure in helium-helium and hydrogen-helium pair correlations 
illustrates the importance of the interaction between the subsystems and is an 
indication that the linear mixing rule is only a first approximation 
for describing mixtures \cite{Vorberger06}.
 
\begin{figure}[t]
\includegraphics[width=.46\textwidth]{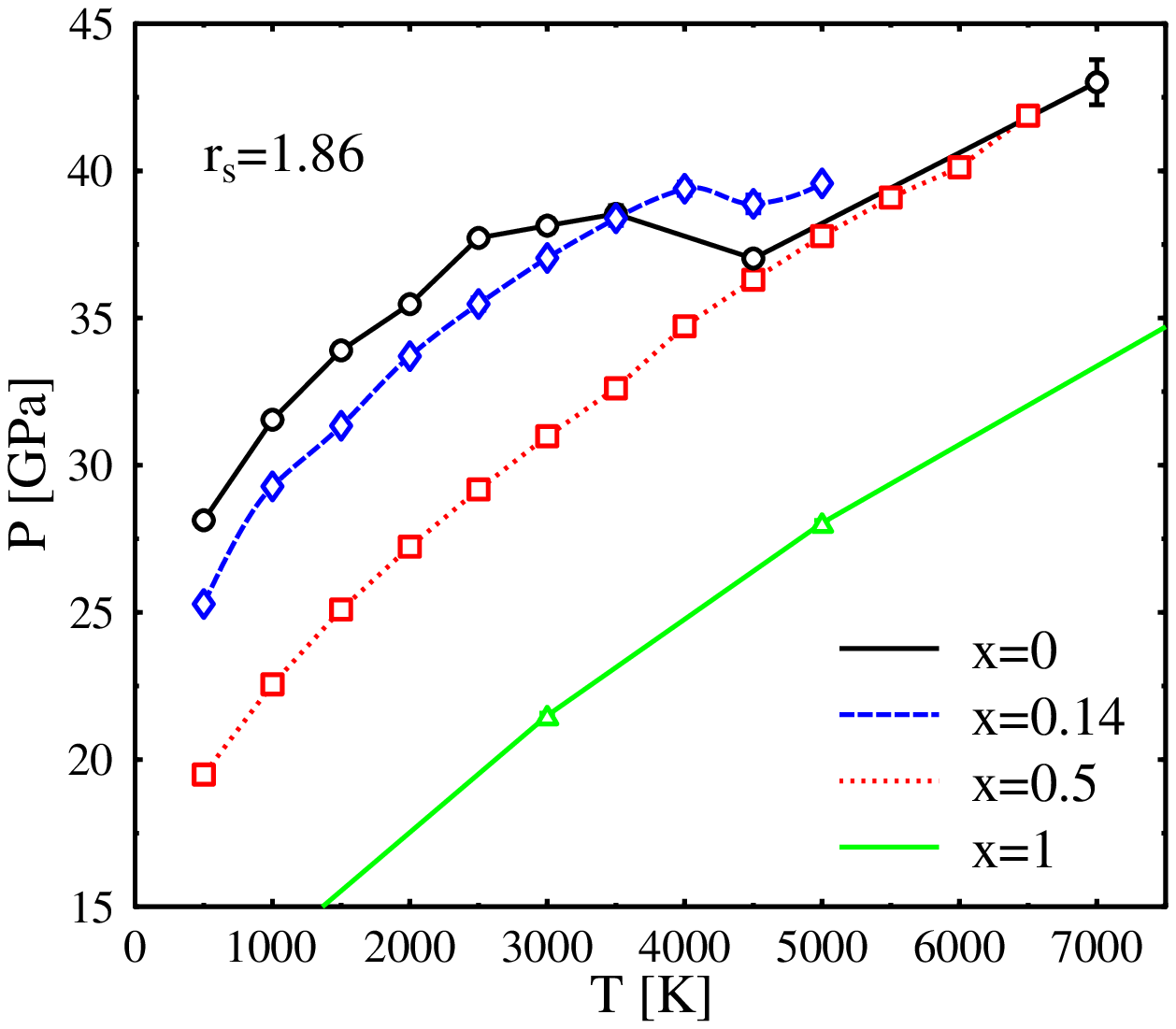}~a)
\hfil
\includegraphics[width=.46\textwidth]{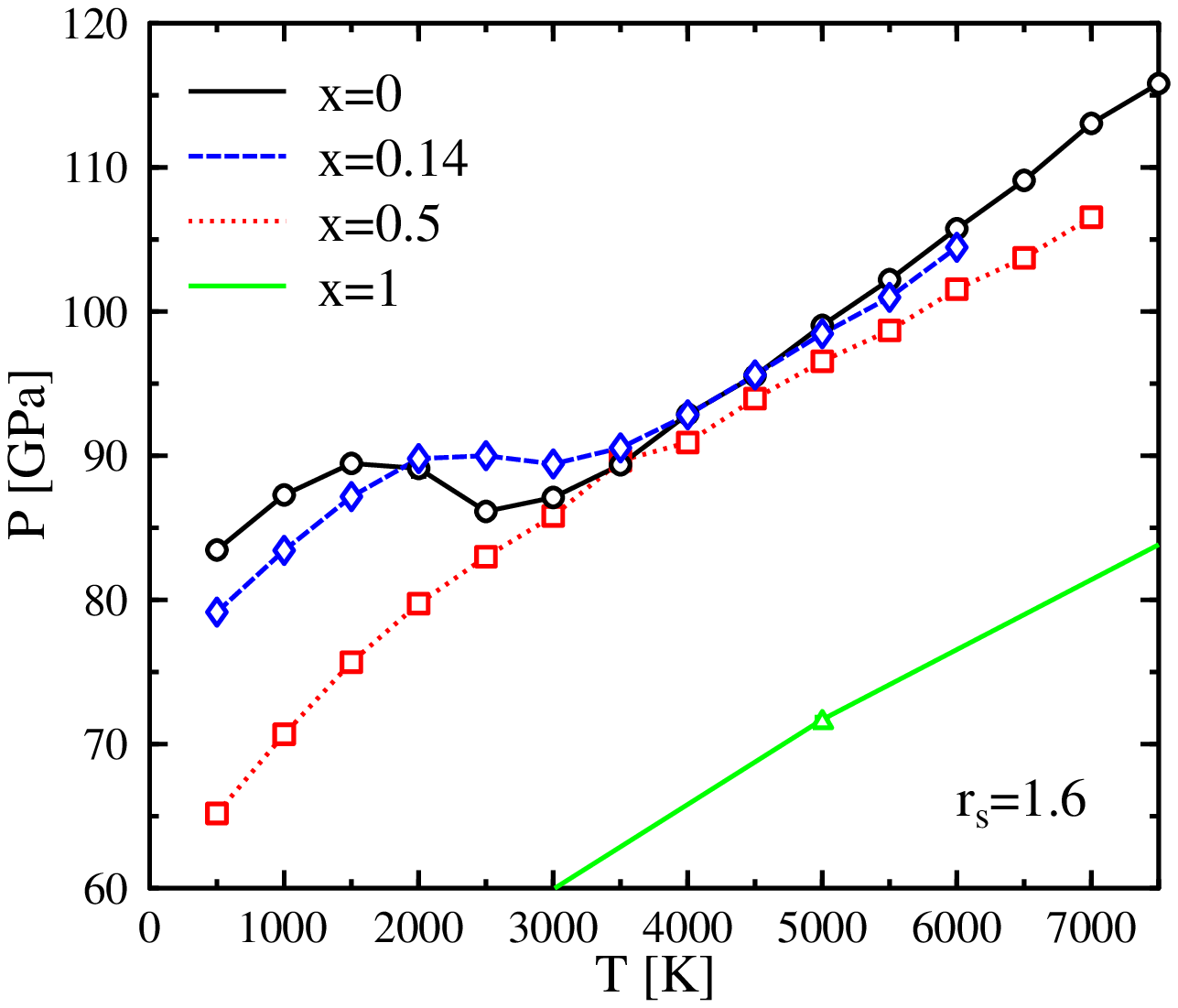}~b)
\caption{Pressure vs. temperature isochores for hydrogen, hydrogen-helium 
mixtures of various mixing 
ratios $x$, and helium for two different electronic density parameters
$r_s$: a) $r_s=1.86$ corresponds to 
$\rho=0.42$ g/cm$^3$ in hydrogen, and b) $r_s=1.6$ corresponds to 
$\rho=0.67$ g/cm$^3$ in hydrogen. 
}
\label{fig:diffx}
\end{figure}
Naturally, 
changes in the structure of the fluid have an effect on the equation of state
(EOS) as presented in Fig. \ref{fig:diffx}. 
The pressure isochores show some distinct features depending on the helium
concentration.
Since helium atoms are smaller than hydrogen molecules, one finds that at
low temperature the pressure decreases substantially when hydrogen molecules are
replaced by helium atoms at constant volume, which leaves the total electronic
density unchanged.
In pure hydrogen, the isochores exhibit a region with a negative 
slope $dP/dT|_V<0$, which is due to the delocalization of the electrons during
the dissociation of hydrogen molecules.
A higher helium fraction reduces the magnitude of the drop in pressure
(e.g. for $x=0.14$) or even completely removes it ($x=0.5$). 
As a result, one finds that the $P(T)$ curves for $x=0$, $x=0.14$, 
and $x=0.5$ approximately converge in the
dissociated regime. There, the pressure depends only little on the helium
concentration because the (ideal) ionic pressure increase and the pressure
decrease due to electronic delocalization depend inversely on the mixing ratio.
Only for large helium concentration this cancellation effect does not occur 
anymore as the comparison with the $x=1$ curve shows.

Fig. \ref{fig:diffx} alone demonstrates temperature 
dissociation. 
By comparing the left and the right part of Fig. \ref{fig:diffx},
it is possible to observe pressure dissociation, too. The region with negative
slopes indicating molecular dissociation is shifted to smaller temperatures 
for higher densities. This means that, by increasing the density at constant
temperature, one can transfer from a molecular fluid to an atomic fluid as well.
%
%
\section{Conclusion}\label{concl}
The structure of the fluid and the equation of state for 
hydrogen-helium mixtures were analyzed
under conditions found in giant gas planets. We have considered the
structural changes in the fluid during dissociation as function of density and helium
concentration. DFT-MD proves to be an useful
technique to study such questions and to obtain accurate equation of state 
data. Such data are needed in order to improve existing models for giant gas 
planets and answer longstanding question concerning their development and 
inner structure.

In particular, we find that helium plays a crucial role in affecting hydrogen 
dissociation both as a function
of temperature and pressure. Hydrogen molecules are stable at higher temperatures 
and pressures in mixtures of hydrogen and helium than in pure hydrogen. The
transition from the molecular to the atomic phase proceeds over a wider range of
temperature and pressure. We observe
regions with negative slope of the pressure isochores 
in pure hydrogen as well as in hydrogen-helium mixtures, but only up to a 
critical mixing ratio.
 
\begin{acknowledgement}
This material is based upon work supported by NASA under the grant NNG05GH29G,
by the Carnegie Institution of Canada, and by the NSF under the
grant 0507321. I.T. and S.A.B. acknowledge support by the NSERC of Canada.
\end{acknowledgement}


\end{document}